\documentclass[12pt,a4paper]{scrartcl}
\usepackage[left=1in,right=1in,top=1in,bottom=1in,a4paper]{geometry}
\usepackage{amsfonts,amsmath,amsthm}
\usepackage{graphicx,dsfont}
\usepackage{pst-plot,pstricks,multido}
\usepackage{enumerate,enumitem}
\usepackage{moreverb}
\usepackage{graphicx,dsfont}
\usepackage{amssymb}
\usepackage{hyperref}
\usepackage{wasysym}
\usepackage{color}

\newtheorem{satz}{Result}[]

\title{Optimal Sample Size Planning for the Wilcoxon-Mann-Whitney-Test}

\author{Martin Happ\thanks{Mail: martin.happ@sbg.ac.at}\\Department of Mathematics,\\University of Salzburg, Austria \and 
	Arne C. Bathke\\Department of Mathematics,\\University of Salzburg, Austria \and
	Edgar Brunner\\Department of Medical Statistics,\\University of G\"ottingen, Germany}

\begin{document}
\maketitle
\section*{Abstract}
There are many different proposed procedures for sample
size planning for the Wilcoxon-Mann-Whitney test at given type-I and type-II
error rates $\alpha$ and $\beta$, respectively. Most methods assume very
specific models or types of data in order to simplify calculations (for
example, ordered categorical or metric data, location shift alternatives,
etc.). We present a unified approach that covers metric data with
and without ties, count data, ordered categorical data, and even dichotomous
data. For that, we calculate the unknown theoretical quantities such as the variances under
the null and relevant alternative hypothesis by considering 
the following ``synthetic data'' approach. 
We evaluate data whose empirical distribution functions match with the
theoretical distribution functions involved in the computations of the unknown
theoretical quantities. Then well-known relations for the ranks of the data are used for the calculations.

In addition to computing the necessary sample size $N$ for a fixed
allocation proportion $t = n_1/N$, where $n_1$ is the sample size in the first
group and $N = n_1 + n_2$ is the total sample size, we provide an interval for
the optimal allocation rate $t$ which minimizes the total sample
size $N$. It turns out that for certain distributions, a balanced design is
optimal. We give a characterization of these distributions. Furthermore
we show that the optimal choice of $t$ depends on the ratio of
the two variances which determine the variance of the Wilcoxon-Mann-Whitney
statistic under the alternative. This is different from an optimal sample size
allocation in case of the normal distribution model.

\section{Introduction} \label{sec:introduction}
The comparison of two independent samples is widespread in medicine, the life 
sciences in general, and other fields of research. Arguably, the most popular 
method is the unpaired $t$-test for two sample comparisons. However, its 
application is limited. For heavy-tailed or very skewed distributions, use of 
the $t$-test is not recommended, especially for small sample sizes. For ordered 
categorical data, comparing averages by means of $t$-tests is not appropriate 
at all. For those situations, a nonparametric test such as the 
Wilcoxon-Mann-Whitney test is much preferred.  

In order to plan a study for such a two sample comparison, we need to know how many subjects are needed to detect a pre-specified effect at least with 
probability $1-\beta$ where $\beta$ denotes the type-II error probability. If 
the underlying distributions are normal, a pre-specified effect might be 
formulated as a difference of means. Within a general nonparametric framework, 
the relative effect (see Section \ref{sec:sample_size}) is very often used. But 
for a statistics practitioner, it is sometimes difficult to state a relevant 
effect size to be detected in terms of the nonparametric relative effect. 
Therefore, we will be using a slightly different approach. Based on prior 
information $F_1$ regarding one group, for example the standard 
treatment or the control group, one can derive the distribution $F_2$ under a 
conjectured (relevant) alternative in cooperation with a subject-matter expert. 
This distribution is established in such a way that it features what the 
subject-matter expert would quantify as a relevant effect. In other words, the 
expert may, but does not necessarily have to, provide a (standardized) 
difference of means as a relevant nonparametric relative effect on which the 
Wilcoxon-Mann-Whitney effect is based. Or alternatively, the subject matter 
expert may simply provide information on an configuration that the expert would 
consider relevant in terms of providing evidence in favour of the research 
hypothesis. This information will then be translated into a relevant 
nonparametric effect. More details on deriving $F_2$ based on an interpretable 
effect in order to compute the nonparametric effect and the variances involved 
in the sample size planning are given in Section \ref{sec:data_examples}. 

For the Wilcoxon-Mann-Whitney test, there already exist many sample size 
formulas. However, most of them require for example either continuous data as 
used in B\"urkner et al. \cite{burkner2017}, Wang et al. \cite{wang2003}, or 
Noether \cite{noether1987}, or they require ordered categorical data as in  
Fan \cite{fan2012}, Tang \cite{tang2011}, Lachin \cite{lachin2011}, Hilton et 
al. \cite{hilton1993}, or Whitehead \cite{whitehead1993}. For a review of 
different methods, we refer to Rahardja et al. \cite{rahardja2009}. A rather 
well known method for sample size calculation in case of continuous data is 
given by Noether \cite{noether1987} who approximated the variance under 
alternative by the variance under the null hypothesis. A similar approximation 
was also used by Zhao et al. \cite{zhao2008} who generalized Noether's formula 
to allow for ties. For practical application however, this approximation may 
not always be appropriate because the variances under null hypothesis and under 
alternative can be very different, thus potentially leading to an under- or 
overpowered study. See, for example, Shieh et al. \cite{shieh2006} for a 
comparison of Noether's formula with different alternative methods.

In some other approaches, the sample size is only calculated under the 
assumption of a proportional odds model for ordered categorical data (e.g. 
Kolassa \cite{kolassa1995} or Whitehead \cite{whitehead1993},  or considering 
only location shift models for continuous metric data (see, e.g., Rosner  and 
Glynn \cite{rosner2009}, Chakraborti et al. \cite{chakraborti2006}, Lesaffre et 
al. \cite{lesaffre1993}, Hamilton \cite{hamilton1991}, or Collings and Hamilton 
\cite{collings1988}, among others). 
An advantage of our Formula (\ref{eq:N}) in Section~\ref{sec:sample_size} for 
the sample size calculation is its generality and practicality. It can be used 
for metric data as well as for ordered categorical data, and it even works very 
well for dichotomous data.
Furthermore, our formula does not assume any special model for the alternatives. 

Within the published literature, the sample size formulas bearing most 
similarity to ours is those by Wang et al. \cite{wang2003}. 
However, their approach is limited to continuous distributions, whereas our 
approach is based on a unified approach allowing for discrete and for 
continuous data.

A completely different way to approach optimality of Wilcoxon-Mann-Whitney 
tests has been pursued by Matsouaka et al. \cite{matsouaka2016}. They use a 
weighted sum of multiple Wilcoxon-Mann-Whitney tests and determine the optimal 
weight for each test. Their aim is not an optimal sample size planning 
including optimization of the ratio of sample sizes, but instead they try to 
optimally combine a primary endpoint with mortality.

In a two sample setting, we sometimes can choose the proportion of subjects in 
the first group. That is, we can choose $t = n_1/N$ where $n_1$ is the number 
of subjects in the first group and $N$ is the total number of subjects. The 
question that arises is how to choose $t$ in an optimal way. In B\"urkner et 
al. \cite{burkner2017}, the optimal $t$ is chosen such that the power of the 
Wilcoxon-Mann-Whitney test is maximized for a given sample size $N$. On the 
other hand, in practice, we prefer to choose $t$ in such a way that the total 
sample size $N$ is minimized for a specified power $1-\beta$. For the two 
sample $t$-test with unequal variances, Dette and O'Brien \cite{dette2004} 
showed that the optimal $t$ to maximize the power of the test is approximately 
\begin{align*}
	t \approx \frac{1}{1 + \tau},
\end{align*}
where $\tau = \sigma_1/\sigma_0$ is the ratio of standard deviations of the two 
groups under the hypothesis and under the alternative, 
respectively. This means that when applying the $t$-test, more subjects should 
be allocated to the group with the higher variance. B\"urkner et al. 
\cite{burkner2017} showed for symmetric, continuous distributions under a 
location shift model, that a balanced design is optimal for the 
Wilcoxon-Mann-Whitney test. For general distributions, they observed in 
simulation studies that in many situations, the difference between using the 
optimal $t$ and using a balanced design is negligible.

In most publications the generation of the alternative from the reference group 
is not discussed and, instead the distribution under the alternative is assumed 
to be known. Here, we want to discuss, however, also how we can generate the 
distribution under the alternative based on the distribution in the reference 
group and an interpretable relevant effect. In order to motivate the method 
derived in this paper, let us consider an example with count data, as it 
appears that most publications on sample size planning focus on ordered 
categorical or continuous metric data. In Table \ref{tab:seizures}, 
the data of an advance information $F_1$ on a placebo for the number of seizures in an epilepsy trial is given. We want to base the sample 
size planning for a new drug on the data $X_{1,1}, \ldots, X_{1,28}$ of the 
advance information $F_1$ which comes from a study published by Leppik et al. 
\cite{leppik1985}, as well as Thall and Vail \cite{thall1990}. For these data, 
we cannot assume a location shift model, as an absolute reduction of two 
seizures would be very good for someone with three seizures, but not really 
helpful for someone with 20 or more seizures. More appropriate would probably 
be a reduction of the number of seizures by some percentage $q$, 
for example $q=50\%$. Based on this specified relevant effect $F_2(x) = 
F_1(x/q)$, we artificially generate a new data set $X_{2,1}, \ldots, X_{2,28}$ 
whose empirical distribution function $\widehat F_2(x)$ is exactly equal to 
$F_2(x)$. Basically, the number $n_2$ of the artificially generated data is 
arbitrary (here, $n_2=28$, e.g.) as long as $\widehat F_2(x) = F_2(x) = 
F_1(x/q)$. We will refer to such data as ``synthetic'' data.

Most of the methods mentioned before cannot be applied to data such as these as 
they have been derived under different, restrictive assumptions. In particular, 
methods assuming a location-shift model cannot be used here. However, 
application of the method proposed in the present paper does not require 
specific types of data or a specific alternative because it is based on the 
observed data and the generated synthetic data, which do not need to follow any 
particular model. See also the Chapter ``Keeping Observed Data as a Theoretical 
Distribution'' in Puntanen et al. \cite{puntanen2011} for a similar approach in 
the parametric case. More details regarding this data set and the sample size 
calculation can be found in Section \ref{sec:data_examples}.  

\begin{table}[ht]
	\centering

	\begin{tabular}{lcrrrrrrrrrrrrrr} \hline
		& & & & & & & & & & \\[-1.5ex]
		\multicolumn{16}{c}{Number of counts} \\[0.4ex] \hline
		& & & & & & & & & & \\[-1.5ex]
		Advance Information & \hspace*{1ex} & \\[0.4ex]
		$X_{1,1} , \ldots, X_{1,28} \sim F_1(x)$ & & 
		3,& 3,& 5, &4, & 21, & 7, & 2, &12, &5, &0, &22, &4, &2, &12 \\
		& & 9, & 5, & 3, & 29, & 5, & 7, & 4, & 4, &5, &8, &25, &1, &2, &12 \\[0.4ex] 
		\cline{3-16}
		& & & & & & & & & & \\[-1.5ex]
		Relevant Alternative & \hspace*{1ex} & \\[0.4ex]
		$X_{2,k} \sim F_2(x) = F_1(x/q)$ & & 
		1, &1, &2, &2, &10, &3, &1, &6, &2, &0, &11, &2, &1, &6 \\
		& & 4, &2, &1, &14, &2, &3, &2, &2, &2, &4, &12, &0, &1, &6 \\[0.4ex] \hline
	\end{tabular} 
	
	\caption{ Number of seizures for 28 subjects from the advance 
		information $X_{1,k} \sim F_1(x)$, $k=1, \ldots, 28$, and for the relevant 
		effect $F_2(x) = F_1(x/q)$, where $q=0.5$ denotes the percentage of the 
		relevant reduction of seizures to be detected. This means $X_{2,k} = [q \cdot 
		X_{1,k}] \sim F_2(x)$, where $[u]$ denotes the largest integer $\le u$.}
	\label{tab:seizures}
\end{table}

The rest of this paper is now organized as follows. We first derive a general 
sample size formula and investigate the behavior of the optimal $t$. That is, 
we show in which cases more subjects should be allocated to the first or second 
group. Then, we apply this method to several data examples with different types 
of data and provide power simulations to show that with the sample size 
calculated by our method, the simulated power is at least $1-\beta$. 
Furthermore, we simulate how the chosen type-I and type-II error rates affect 
the value of the optimal allocation rate $t$.


\section{Sample Size Formula}
\label{sec:sample_size}
Let $X_{1i} \sim F_i$ and $X_{2j} \sim F_2$, $i = 1, \dots, n_1$, $j = 1 \dots, n_2$,
be independent random samples obtained on $N$ different subjects, with $N = n_1 + n_2$.
The cumulative distribution functions $F_1$ and $F_2$ are understood as their normalized versions, 
that is $F_i(x) = \tfrac{1}{2}\big(F_i^+(x) + F_i^-(x)\big)$ where $F_i^+$ denotes the right-continuous, 
and $F_i^-$ denotes the left continuous cumulative distribution function. 
By using the normalized version, we can pursue a unified approach for continuous and discrete data,
no separate formulas ``correcting for ties'' are necessary. 
This unified approach results naturally in the usage of midranks in the formulas for the test statistics, see Ruymgaart 
\cite{ruymgaart1980}, Akritas, Arnold and Brunner \cite{akritas1997}, and Akritas and Brunner \cite{akritas1997b}  for details.
With $t$, we denote the proportion of the $N$ subjects that is allocated to the first group.
That is, $n_1 = tN$ and $n_2 = (1-t)N$. 
Without loss of generality, $X_{1i}$ may be regarded as the reference group, 
and the second group $X_{2i}$ as the (experimental) treatment group. 
The Wilcoxon-Mann-Whitney test is based on the nonparametric relative treatment effect 
\begin{align}
	p &= \int F_1 d F_2 
	= P(X_{11} < X_{21} ) + \frac{1}{2} P( X_{11} = X_{22}) \label{ssf:releff}
\end{align}
which can be estimated in a natural way by its empirical analog $\hat{p} = \int \hat{F}_1 d\hat{F}_2$.
Here, $\hat{F}_i = \frac12 (\hat{F}^-_i + \hat{F}^+_i )$ is the normalized empirical cumulative distribution function with $\hat{F}^-_i(x) = n_i^{-1} \sum_{j=1}^{n_i} \mathds{1}_{\{ X_{ij}< x\}}$,
and $\hat{F}^+_i(x) = n_i^{-1} \sum_{j=1}^{n_i} \mathds{1}_{\{ X_{ij}\leq x\}}$ the left and right continuous empirical cumulative distribution functions for $i = 1,2$, respectively.
Finally, $\mathds{1}_{\{ X_{ij}< x\}}$ denotes the indicator function of the set ${\{ X_{ij}< x\}}$. 
Using the asymptotic equivalence theorem, see for example Brunner and Munzel \cite{brunner2000} or Brunner and Puri \cite{brunner2001}, it can be shown that the statistic
\begin{align}
	T_N = {\sqrt{N}(\hat{p} - p) },
\end{align}
is asymptotically normal under slight regularity assumptions. Let us denote by
\begin{align} \label{eq:equivalence}
	U_N = \sqrt{N}\Big( n_2^{-1} \sum_{j=1}^{n_2} F_1(X_{2j}) -  n_1^{-1} \sum_{j=1}^{n_1} F_2(X_{1j})  + 1 - 2p  \Big)
\end{align}
the statistic that is an asymptotically equivalent statistic to $T_N$, but based on independent random variables. 
Then, under the null hypothesis $H_0: F_1 = F_2$, the variance of $U_N$ can be written as
\begin{align} \label{eq:variance_null}
	\sigma_0^2 = \frac{N^2}{n_1 n_2}\sigma^2 = \frac{1}{t(1-t)}\sigma^2,
\end{align}
where $\sigma^2 = \int F_1^2 dF_1 - \tfrac{1}{4}$.
This means, $T_N/\sigma_0$ has asymptotically the same distribution as $U_N/\sigma_0$, but the distribution of the latter is asymptotically standard normal.
To compute the variance of $T_N$ under the alternative hypothesis, 
we again take advantage of this asymptotic equivalence in (\ref{eq:equivalence}) and obtain the following  asymptotic 
variance $\sigma_N^2$ under alternative.
\begin{align}
	\sigma_N^2 = \frac{N}{n_1 n_2}(n_2 \sigma_1^2 + n_1 \sigma_2^2) \label{ssf:sigmanq}
\end{align}
where 
\begin{align}
	\sigma_1^2 &= Var(F_2(X_{11})) = \int F_2^2 d F_1 - (1-p)^2,   \label{ssf:sigma1q} \\
	\sigma_2^2 &= Var(F_1(X_{21})) = \int F_1^2 d F_2 - p^2. \label{ssf:sigma2q}
\end{align}

Clearly, the variance $\sigma_N^2$ under alternative is a weighted sum of two components, $\sigma_1^2$ and $\sigma_2^2$. 
Both of these components are important for minimizing the sample size, as performed in Section \ref{sec:minimizing}, unlike the parametric case where only the two variances $\sigma_0^2$ under the null and $\sigma_1^2$ under the alternative hypotheses are considered.

Based on these considerations, an approximate sample size formula for the Wilcoxon-Mann-Whitney test can be obtained similar to the one calculated by Wang et al. \cite{wang2003} for continuous data. 
Namely, we obtain 
\begin{align} \label{eq:N_general}
	N = \frac{\Big( \sigma_0 u_{1-\alpha/2} + \sigma_N u_{1-\beta}  \Big)^2}{(p-\frac{1}{2})^2},
\end{align}
where $\alpha$ and $\beta$ denote the type-I and type-II error rates, respectively, 
and $u_{1-\alpha/2}$ is the $1-\alpha/2$ quantile of the standard normal distribution. 

The quantities $p, \sigma_0$, and $\sigma_N$ in Equation (\ref{eq:N_general}) are unknown in general. Moreover, $\sigma_N^2$ is a linear combination of the two unknown variances $\sigma_1^2$ and $\sigma_2^2$ in Equations~(\ref{ssf:sigma1q}) and (\ref{ssf:sigma2q}).
To compute these quantities from the distribution $F_1$ of the prior information in the reference group and the distribution $F_2$ 
generated by an intuitive and easy to interpret relevant effect, we proceed as follows.

We interpret the distributions of the data as fixed theoretical distributions similar to the parametric case in Seber \cite{seber2008} on page 433 and Puntanen et al. \cite{puntanen2011} on pages 27 and 28. Therefore, we denote the data from the prior information by 
$X_{11}^*, \ldots, X_{1n_1}^*$ and the synthetic data for the treatment group by $X_{21}^*, \ldots, X_{2n_2}^*$. The corresponding cumulative distribution functions are denoted by $F_1^*(x) = \hat{F}_1(x)$ and $F_2^*(x) = \hat{F}_2(x)$, respectively. Here, 
$\hat{F}_1(x)$ denotes the empirical distribution function of the available data $X_{11}^*, \ldots, X_{1n_1}^*$ in 
the reference group and $\hat{F}_2(x)$ the empirical distribution functions of the synthetic data $X_{21}^*, \ldots, X_{2n_2}^*$ in 
the treatment group. In this context, ``synthetic'' means that the data for $F_2$ are artificially generated based on the prior 
information $F_1$ and some interpretable relevant effect. We can generate data sets of arbitrary size for $F_1$ and $F_2$, as long as 
the relative frequencies or probabilities remain unchanged. Because we assume that our synthetic data represent fixed distributions and
not a sample, we can calculate the variances $\sigma_1^2$, $\sigma_2^2$, and $\sigma^2$, as well as the relative effect $p$ \textit{exactly}. To emphasis that these quantities are not estimators but rather the \textit{true parameters} based on the \textit{synthetic data}, 
we will denote these quantities by $\sigma^{2*}, \sigma_1^{2*}, \sigma_2^{2*}$, and $p^*$.  

By using the relations $N t= n_1$ and $N(1-t) = n_2$, the sample size formula from Equation (\ref{eq:N_general}) is then rewritten as 
\begin{align} \label{eq:N}
	N = \frac{\Big( \sigma^* u_{1-\alpha/2} + u_{1-\beta} \sqrt{t \sigma_2^{2*} +(1-t) \sigma_1^{2*} }  \Big)^2}{t(1-t)(p^*-\frac{1}{2})^2}.
\end{align}

The variances and the relative effect can be easily calculated by using a simple relation between ranks and the so-called placements
$P_{1k} = n_2 \hat{F}_2(X_{1k})$ and $P_{2k} = n_1 \hat{F}_1(X_{2k})$, which were introduced by Orban and Wolfe \cite{orban198, 
	orban1980}. The placements were first defined only for continuous distributions, but were later generalized to include discrete 
distributions, for details see, for example, Brunner and Munzel~\cite{brunner2000}. To this end, let $R_{ik}^*$ denote the overall rank of 
$X_{ik}^*$ among all $n_1 + n_2 = N$ synthetic data, and $R_{ik}^{*(i)}$ the ranks within the $i$-th group, $i = 1,2$. Further, let 
$\overline{R}_{i\cdot}^{\ *} = \frac{1}{n_i} \sum_{k = 1}^{n_i} R_{ik}^*$, $i = 1,2$, denote  the rank means. Then, the placements 
$P_{ik}^*$ can be represented by these ranks as $P_{ik}^* = R_{ik}^*- R_{ik}^{*(i)}$, $i = 1,2; k=1, \ldots, n_i$. Finally, by letting 
$F_i^*(x) = \hat{F}_i(x)$, the quantities in the sample size formula (\ref{eq:N}) can be calculated directly as follows.
\begin{align}
	p^* &= \int F_1^* dF_2^*  =   \frac{1}{N} (\overline{R}_{2\cdot}^{\ *} - \overline{R}_{1\cdot}^{\ *} ) + \frac{1}{2},\\
	\sigma^{2*} &= \int F^{2*} dF^* - \frac{1}{4} =  \frac{1}{N^3} \sum_{i=1}^{2} \sum_{k = 1}^{n_i} \Big( R_{ik}^* - \frac{N + 1}{2} 
	\Big)^2,\\
	\sigma_1^{2*} &= \int F_2^{2*} dF_1^{*} - (1-p^*)^{2} = \frac{1}{n_1 n_2^2} \sum_{k = 1}^{n_1} \Big( P_{1k}^* - 
	\overline{P}_{1 \cdot}^{\ *}  \Big)^2, \label{eq:sig1} \\
	\sigma_2^{2*} &= \int F_1^{2*} dF_2^{*} - {p^*}^{2} = \frac{1}{n_1^2 n_2} \sum_{k = 1}^{n_2} \Big( P_{2k}^* - 
	\overline{P}_{2\cdot}^{\ *}  \Big)^2. \label{eq:sig2}
\end{align}

Note that for computing the variances, we do not divide by $N-1$ or $n_i - 1$, but rather by $N$ or $n_i$, $i = 1, 2$ because the 
distributions of the synthetic data are considered as fixed theoretical distributions similar to the parametric case in Puntanen et 
al. \cite{puntanen2011} (pages 27 and 28).

\section{Minimizing $N$}
\label{sec:minimizing}
\subsection{Interval for the optimal design}
\label{sec:interval}
In Section \ref{sec:sample_size} we have derived a formula for the sample size $N$ given type-I and type-II error rates $\alpha$ and $\beta$, respectively. 
In practice, we sometimes have the opportunity to choose how many subjects should be allocated to the first group and how many to the second. 
The question in such a situation is how the proportion $t = n_1/N$ should be chosen in order to minimize $N$.  B\"urkner et al. \cite{burkner2017} aimed at finding the optimal $t$ such that the power is maximized for a given sample size $N$. 
Although both questions lead to essentially the same answer, we prefer to minimize the sample size as this question arises more naturally in sample size planning.

Technically, an exact solution to this problem is possible, but it is not feasible to write down the solution in closed form anymore, and it does not give us much information about the behaviour of the solution. However, it is possible to provide an interpretable interval for the optimal allocation rate $t_0 = \arg\min_{t \in (0,1 ) } N(t)$. For that, we only have to assume that the power $1-\beta$ is greater than $50$ \% and we distinguish between the cases $\sigma_1 = \sigma_2$ and $\sigma_1 \neq \sigma_2$. Note that the variances $\sigma_1^{2}$ and $\sigma_2^{2}$ can be quite different even if the variances of  $F_1$ and $F_2$ are the same. If we allow unequal variances for $F_1$ and $F_2$ it is even possible that $\sigma_1^{2} = 0$ and $\sigma_2^{2} = 1/4$ occurs where $1/4$ is the largest possible value for the variances $\sigma_i^{2}$, $i=1,2$.

The assumption on the minimal power could be weakened to assuming that the numerator of $N(t)$ is not zero. One then only needs to distinguish the cases $\beta > 1/2$, $\beta < 1/2$, and $\beta = 1/2$. 
For practical considerations, however, only $\beta < 1/2$ is of relevance, therefore we only consider this situation.

Now regarding the case  $\sigma_1 = \sigma_2$, it is clear from Formula (\ref{eq:N}) that the optimal allocation rate is $t_0 = 1/2$ because the numerator of $N(t)$ does not depend on $t$ and $t(1-t)$ is maximized at $t=1/2$. 
For the case $\sigma_1 \neq \sigma_2$ we consider first  $0 < \sigma_1 < \sigma_2$. 
Then it is possible to show (see Appendix, Result \ref{app:result2}) that
the sample size is minimized by a $t_0 \in [I_1, I_2]$ with $I_1 \leq I_2 < 1/2$. The minimizer is unique in the interval $(0,1)$ and the bounds $I_1$ and $I_2$ are given by
\begin{align} 
	I_1 &= \frac{1}{\kappa + 1}, \label{eq:I1} \\
	I_2 &= \frac{\sqrt{z}}{ \sqrt{z} + (u_{1-\alpha/2}~ \sqrt{q} \sigma + u_{1-\beta}~ \sigma_2^{2} )  }, \label{eq:I2}
\end{align}
where $\kappa = \sigma_2/\sigma_1$,~ $\sigma^2 = \int F_1^2 dF_1 - 1/4 $ as in (\ref{eq:variance_null}),~ $q = p(1-p)$, 
and
\begin{align*} 
	z = \left( u_{1-\alpha/2}~\sqrt{q} \sigma + u_{1-\beta}~\sigma_1^2 \right) \left(u_{1-\alpha/2}~\sqrt{q} \sigma + 
	u_{1-\beta}~\sigma_2^2 \right).
\end{align*}

Additionally, the following equivalence holds
\begin{align} 
	t_0 < \tfrac{1}{2} &\Longleftrightarrow \sigma_1 <  \sigma_2.
\end{align} 

In the case $0 < \sigma_2 < \sigma_1$, we obtain an analogous result for the minimizer $t_0 \in [I_2, I_1]$, where the bounds are the same as before. 
Moreover we have a similar equivalence, namely
\begin{align}
	t_0 > \tfrac{1}{2} &\Longleftrightarrow \sigma_1 >  \sigma_2.
\end{align}
The derivation of these two equivalences can be found in the Appendix in the Results \ref{app:result2} and \ref{app:result3}.

From the form of the interval $[I_1, I_2]$ we can see that if  $\kappa \approx 1$ then $t_0 \approx 1/2$. 
In most cases this means that the minimum total sample size $N$ is obtained for allocation rates 
close to $1/2$, or the allocation rate is $1/2$ because of rounding. 
Larger values for the type-I error rate $\alpha$ or the power $1-\beta$ lead in general to more extreme values for $t_0$, that is $|1/2 - t_0|$ gets larger. This can be seen from the upper bound $I_2$. By increasing $\alpha$ or the power $1-\beta$ the bound $I_2$ decreases (or increases for $\sigma_1 >  \sigma_2$). Typically this means that the difference $|1/2-t_0|$ tends to get larger. Note that $I_2$ is bounded from below (above), that is $t_0$ cannot become arbitrarily small (or large). 
The impact of $\alpha$ and $\beta$ is demonstrated in simulations in Section \ref{sec:simulations}.  

Next, we consider the case  $0 = \sigma_1 < \sigma_2$.  In the same way as before, it is possible to construct an interval for the optimal allocation rate $t_0$ which is given by $[I_1^{(0)}, I_2]$, where the lower bound is
\begin{align}
	I_1^{(0)} = \frac{u_{1-\alpha/2}~ \sigma }{2 u_{1-\alpha/2}~ \sigma + u_{1-\beta}~ \sigma_2 },
\end{align}
and the upper bound is the same as in the case $0 < \sigma_1$. 
More details are given in the Appendix in Result \ref{app:result4}. 
An analogous result can be obtained for $0 = \sigma_2 < \sigma_1$.

Therefore, the value of $t_0$ is mainly determined by $\kappa$ which is the ratio of the standard deviations $\sigma_1$ and $\sigma_2$ under the alternative hypothesis. 
This is qualitatively different from the result of Dette and O'Brien \cite{dette2004} for the $t$-test in a parametric location-scale model, 
where the optimal allocation value is determined by the ratio of standard deviations under the null and under the alternative hypothesis.
For the Wilcoxon-Mann-Whitney test, the variance under null hypothesis is not really important for determining $t_0$, in case of continuous distributions, for example, the variance under null hypothesis is $\sigma_0^2 = 1/12$.

\subsection{Optimality of a Balanced Design}
In the previous section, we have provided ranges for the optimal allocation proportion $t_0$.
There are many situations, in which balanced designs are optimal or close to optimal.
In this section, we will describe classes of situations in which a balanced design minimizes the sample size. 
From Section \ref{sec:interval} we know that 
\begin{align}
	t_0 = \frac{1}{2} \Longleftrightarrow \sigma_1 = \sigma_2.
\end{align}
The right hand side of this equivalence can be rewritten as
\begin{align}
	t_0 = \frac{1}{2} \Longleftrightarrow \sigma_1 = \sigma_2 \Longleftrightarrow \int F_1^{2} dF_2 = \int (1-F_2)^2 dF_1.
\end{align}
B\"urkner et al. \cite{burkner2017} showed analytically that for symmetric and continuous distributions with $F_2(x) = F_1(x + a)$ and $a \neq 0$, the minimal sample size is attained at $t_0 = 1/2$. 
Such distributions satisfy the integral equation
\begin{align}\label{eq:integrals}
	\int F_1^2 dF_2 = \int (1-F_2)^2 dF_1.
\end{align}
But the class of distributions satisfying Equation (\ref{eq:integrals}) is actually larger. 
Consider normalized cumulative distribution functions $F_1, F_2$ for which an $a \in \mathbb{R}$ exists such that for all $x \in \mathbb{R}$ the following equality holds
\begin{align}\label{eq:distributions}
	F_1(a + x ) = 1 - F_2(a - x) \ .
\end{align}

Further, let us assume $1-\beta > 0.5$. Then, the minimum for $N(t)$, $t \in (0,1)$, is attained at $t_0 = 1/2$. This 
means that (\ref{eq:distributions}) is a sufficient but not necessary condition for $t_0 = 1/2$. As an example for distributions that satisfy Equation (\ref{eq:integrals}) but not (\ref{eq:distributions}) consider $F_1 = F_2$ to be a non-symmetric distribution.

Note that we do not assume for (\ref{eq:distributions}) that the distributions are stochastically ordered or symmetric. 
If we assume finite third moments then equation (\ref{eq:distributions}) only implies that both distributions have the 
same variance and their skewness has opposite signs, that is, $\nu_{F_1} = - \nu_{F_2}$ if we denote with $\nu_{F_i}$ the skewness of
the distribution with cdf $F_i$, $i = 1,2$. 

Obviously, for a large class of distributions, the optimal allocation rate is exactly $1/2$. 
B\"urkner et al. \cite{burkner2017} already noticed the robustness of the Wilcoxon-Mann-Whitney test regarding the optimal allocation rate. 
When the optimal $t_0$ is not equal to $1/2$, it is often close to $1/2$.
Furthermore, the exact choice of $t$ typically only has a small influence on the required total sample size. 
This applies not only to continuous and symmetric distributions but in general to arbitrary distributions. 


\section{Data Examples}
\label{sec:data_examples}
The generality of the approach proposed in this paper is demonstrated using different data examples with continuous metric, discrete metric, and ordered categorical data. 
In this section, we first describe the data sets. 
Then, the calculated sample sizes along with the actual achieved power in comparison with other sample size calculation methods are given. 
For all data sets, we used the prior information from one group (e.g., from a previous study or from literature) to generate synthetic data for the second group based on an interpretable effect specified by a subject matter expert. 
For ordered categorical data, such an effect might be that a certain percentage of subjects in each category are moved to a better or worse category. 
For metric data, it is possible to simply use a location shift as the effect of interest. 
Regardless on how the effects are chosen, in the end, they all are translated into the so-called nonparametric relative effect which itself provides for another interpretable effect quantification which might be useful for practitioners, in addtion to, for example, a location shift effect.

\label{sec:data_results}
For all examples, we used $\alpha = 0.05$ as the type-I error rate and provide the output from an R function which shows the optimal $t$, the sample size determined for each group, 
and the ratio $\kappa = \sigma_2/\sigma_1$. 
Furthermore, we provide simulation results to assess the actual achieved power. 
The R Code is given in the appendix. 
For calculating the asymptotic Wilcoxon-Mann-Whitney test, we used the function \verb|rank.two.samples| from the R package rankFD 
\cite{rankFD}. 
For all simulations performed with the statistical software R, we generated $10^4$ data sets and used $0$ as our starting seed value for drawing data sets from the synthetic data. 
To compute the optimal allocation rate $t_0$ and the sample sizes for each group, the function \verb|WMWssp_Minimize| from the R package rankFD \cite{rankFD} can be used.

\subsection{Number of Seizures in an Epilepsy Trial}
The data for the placebo group of a clinical trial published in Thall and Vail \cite{thall1990} and Leppik et al. \cite{leppik1985} is shown in Table \ref{tab:seizures}. 
As mentioned in the Introduction, a relevant effect for a drug may be stated as a reduction of the number of seizures by $50\%$. 
A location-shift model is clearly not appropriate for these data. 
Based on the specified relevant effect size, we can generate synthetic data.
They lead to a nonparametric relative effect $p$ of approximately $0.27$ which is inserted into the sample size formula. 

In order to have a power of at least $80\%$, we need $24$ subjects in each group,
according to our method. 
When using the optimal $t_0 \approx 0.49$, we need $n_1 = 23$ and $n_2 = 24$ subjects.
In this case, the optimal allocation only reduces the total number of subjects needed by one, 
in comparison with a balanced design. 
Applying Noether's formula  in this case yields sample sizes $n_1 = n_2 = 26$. 
Table \ref{tab:power_seizures} presents results from a power simulation regarding the different sample size recommendations. 
Here, Noether's formula would lead to a slightly overpowered study. 
\begin{table}[ht]
	\centering
	\begin{tabular}{lcccccl} 	\hline
		Method & \hspace*{5ex}     & Sample Sizes $n_1/n_2$ & \hspace*{3ex} & Total Sample Size $N$  & \hspace*{3ex} & Power \\ \hline
		Balanced & \hspace*{5ex}   & 24/24 & \hspace*{3ex} & 48 & \hspace*{3ex} & 0.802 \\
		Unbalanced & \hspace*{5ex} & 23/24 & \hspace*{3ex} & 47 & \hspace*{3ex} & 0.7956 \\
		Noether & \hspace*{5ex}    & 26/26 & \hspace*{3ex} & 52 & \hspace*{3ex} & 0.8417 \\ \hline
	\end{tabular} 
	\caption{Power simulation for the number of seizures.} 	\label{tab:power_seizures}
\end{table} 

\subsection{Irritation of the Nasal Mucosa}
In this study, two inhalable substances with different concentrations are compared with regard to the severity of the nasal mucosa 
damage of rats (see Akritas, Arnold and Brunner \cite{akritas1997}). The severity of irritation is described using a defect score from 
$0$ to $3$ where $0$ refers to no irritation and $3$ to severe irritation. For the nasal mucosa data, we have prior information for 
substance~1 with 2~ppm concentration. A pathologist suggests, for example, that a worsening of one score unit for $25\%$ 
of the rats in categories 0, 1, and 2 is a relevant effect. This means that $25\%$ of the rats with score $0$ will be assigned 
score~$1$ and so forth. The resulting synthetic data set for substance~2 is given in Table \ref{tab:nasal_mucosa}. 
The original data set for substance 1 has been augmented by factor~$4$ in order to obtain integer values of the samples sizes for 
the synthetic data for substance~2. The result of the sample size calculation is not affected by this because the relative frequencies 
for substance~1 remain unchanged.
\begin{table}[ht]
	\centering
	\begin{tabular}{lccccc} 
		& \hspace*{3ex} & \multicolumn{4}{c}{Defect Score} \\ \hline	
		& \hspace*{3ex} & 0 & 1 & 2 & 3 \\ \hline		
		Substance 1 & \hspace*{3ex} & 64 & 12 & 4 & 0 \\
		Substance 2 & \hspace*{3ex} & 48 & 25 & 6 & 1 \\ \hline
	\end{tabular}
	\caption{Number of rats with defect score $0, 1, 2, 3$.} \label{tab:nasal_mucosa}
\end{table}

Based on the synthetic data in Table \ref{tab:nasal_mucosa}, the relative effect is $p = 0.599$. 
Performing a sample size calculation with $1-\beta = 0.8$ and balanced groups results in sample sizes $n_1 = n_2 = 85$. For this data set, the ratio of variances $\kappa$ is larger than $1$, therefore it is beneficial to assign fewer subjects to the first group (substance 1). To be more precise, the optimal allocation rate $t_0$ is approximately $0.49$ which leads to sample sizes $n_1 = 83$ and $n_2 = 87$. 
But as we can see, in both cases the total sample size is $N = 170$. 
If we apply Noether's formula \cite{noether1987}, we arrive at $n_1 = n_2 = 134$ which is considerably larger than the estimated minimal sample size based on our method 
and leads to a remarkably overpowered study, with actual power of over $94\%$ (see Table {\ref{tab:power_nasal_mucosa} for the simulation results). 
	This is mainly due to ties in the data. 
	Recall that Noether's formula was derived for continuous distributions. 
	Our method achieves $80\%$ power for the balanced and unbalanced design. 
	Tang \cite{tang2011} derived a sample size formula for ordered categorical data. 
	If we use his method, we obtain that 86 rats per group are needed. 
	The closeness of his result to ours may be taken as confirmation that our unified approach
	produces appropriate results also in the case of ordered categorical data.

	\begin{table}[ht]
		\centering
		\begin{tabular}{lcccccl} \hline
			Method & \hspace*{3ex} & Sample Sizes $n_1/n_2$ & \hspace*{3ex} & Total Sample Size $N$ & \hspace*{3ex} & Power \\ 	\hline
			Balanced & \hspace*{3ex} & 85/85 & \hspace*{3ex} & 170 & \hspace*{3ex} & 0.8027 \\
			Unbalanced & \hspace*{3ex} & 83/87 & \hspace*{3ex} & 170 & \hspace*{3ex} & 0.7999 \\
			Noether & \hspace*{3ex} & 134/134 & \hspace*{3ex} & 268 & \hspace*{3ex} & 0.9417 \\
			Tang & \hspace*{3ex} & 86/86 & \hspace*{3ex} & 172 & \hspace*{3ex} & 0.8045 \\ \hline
		\end{tabular}
		\caption{Power simulation for the nasal mucosa data.} \label{tab:power_nasal_mucosa}
	\end{table}
}

\subsection{Kidney Weights}
In this placebo-controlled toxicity trial, female and male Wistar rats have been given a drug in four different dose levels. 
The primary outcome is the relative kidney weight in [\permil], that is the sum of the two kidney weights divided by 
the total body weight, and multiplied by 1,000. For calculating the sample size we consider only male rats from the placebo group and 
generate a suitable data set exhibiting a relevant effect for the treatment group.  
For generating the synthetic data of the treatment group, 
an expert considers a location shift of $5\%$ of the mean from the placebo group as a relevant effect. 
The data are displayed in Table \ref{tab:kidney}.
\begin{table}[hb]
	\centering
	%
	\begin{tabular}{lcllllllll}
		& \hspace*{3ex} & \multicolumn{8}{c}{Relative Kidney Weight [\permil]} \\ \hline
		Placebo & \hspace*{3ex} & 6.62 & 6.65  & 5.78 & 5.63 & 6.05 & 6.48 & 5.50 & 5.37  \\ 
		Treatment & \hspace*{3ex} & 6.92 & 6.95 & 6.08 & 5.93 & 6.35 & 6.78 & 5.80 & 5.67 \\ 
		\hline
	\end{tabular}	
	\caption{Relative kidney weights [\permil] for 16 male Wistar rats.} \label{tab:kidney}	
\end{table}

Using the data from Table \ref{tab:kidney} as our synthetic data, 
the nonparametric relative effect is calculated as $p \approx 0.70$.
Thus, we need $n_1 = n_2 = 30$ Wistar rats to have a power of at least $80\%$. 
In this example, there is again barely any difference between using the optimal design $t_0 \approx 0.51$ ($n_1 = 31$, $n_2 = 30$) 
and a balanced allocation. Because of rounding, in this case the optimal design even leads to a larger sample size $N = 61$ in 
comparison to $N = 60$ obtained using a balanced design. 
Noether's formula leads to sample sizes $n_1 = n_2 = 32$ in this case. 
The simulated power is given in Table \ref{tab:power_kidney}. 
Clearly, Noether's formula again exceeds the $80\%$ power. 
Our method maintains the power quite well and leads to just a slight inflation 
of power in the unbalanced design.
\begin{table}[ht]
	\centering
	\begin{tabular}{lcccccl}
		\hline
		Method & \hspace*{5ex} & Sample Sizes $n_1/n_2$ & \hspace*{3ex} & Total Sample Size $N$ & \hspace*{3ex} & Power \\ \hline
		Balanced & \hspace*{5ex} & 30/30 & \hspace*{3ex} & $60$ & \hspace*{3ex} & 0.7976 \\
		Unbalanced & \hspace*{5ex} & 31/30 & \hspace*{3ex} & $61$ & \hspace*{3ex} & 0.8123 \\
		Noether & \hspace*{5ex} & 32/32 & \hspace*{3ex} & $64$ & \hspace*{3ex} & 0.8320 \\ \hline
	\end{tabular}
	\caption{Power simulation for the relative kidney weights.} \label{tab:power_kidney}
\end{table} 

\subsection{Albumin in Urine}
This data set was considered by Lachin \cite{lachin2011} and contains albumin levels in the urine (albuminuria) of diabetic patients. The levels of albumin are rated as either normal, microalbuminuria, or macroalbuminuria. 
The goal of the study was to compare two treatments, with expected conditional probabilities as given in Table \ref{tab:albumin}.
\begin{table}[ht]
	\centering	
	\begin{tabular}{lclclcl} \hline
		& \hspace*{5ex} & Normal & \hspace*{3ex} & Micro & \hspace*{3ex} & Macro \\ \hline
		Control & \hspace*{5ex} & 0.85 & \hspace*{3ex} & 0.10 & \hspace*{3ex} & 0.05 \\ 
		Experimental & \hspace*{5ex} & 0.90 & \hspace*{3ex} & 0.075 & \hspace*{3ex} & 0.025 \\ \hline
	\end{tabular}
	\caption{Relative frequencies for the Albumin data from Lachin \cite{lachin2011}.} \label{tab:albumin}
\end{table}

For $90\%$ power, Lachin \cite{lachin2011} reports a required sample size of $N = 1757$ ($1758$ because of rounding to achieve balanced sample sizes). 
Using our proposed method, we obtain a necessary total sample size of $N = 1754$ in the balanced case. 
For the optimal design, we obtain $N = 1751$ with an optimal allocation rate $t_0$ around $0.52$. 
Simply using the Noether formula despite the ties, one would calculate a required sample size of
$N = 5334$ (!), clearly leading to a much overpowered study. 
The other three methods attain the nominal power based on a simulation study.
The relative effect for this data set is $p = 0.474$.
\begin{table}[ht]
	\centering	
	\begin{tabular}{lcccccl}
		\hline
		Method & \hspace*{5ex} & Sample Sizes $n_1/n_2$ & \hspace*{3ex} & Total Sample Size $N$ & \hspace*{3ex} & Power \\ \hline
		Balanced & \hspace*{5ex} & 877/877 & \hspace*{3ex} & 1754 & \hspace*{3ex} & 0.9054 \\
		Unbalanced & \hspace*{5ex} & 909/842 & \hspace*{3ex} & 1751 & \hspace*{3ex} & 0.9033 \\
		Lachin & \hspace*{5ex} & 879/879 & \hspace*{3ex} & 1758 & \hspace*{3ex} & 0.9029 \\
		Noether & \hspace*{5ex} & 2667/2667 & \hspace*{3ex} & 5334 & \hspace*{3ex} & $\approx 1$ \\
		\hline
	\end{tabular}
	\caption{Power simulation for the albumin in urine data.} \label{tab:power_albumin}
\end{table}

In the above four data examples, we have used $\alpha = 0.05$ and $1-\beta = 0.8$ or $0.9$ for the sample size calculation and power simulation. 
According to Formula (\ref{eq:N}) and the intervals for $t_0$ 
(Equations (\ref{eq:I1}) and (\ref{eq:I2}) in Section \ref{sec:interval}), 
the choice of $\alpha$ and $\beta$ has an influence not only on the total sample size $N$,
but also on the optimal allocation rate $t_0$. 
In order to study the behaviour of these two parameters, we have performed two simulation studies which are described in 
Section~\ref{sec:simulations}.

\section{Simulations for the Optimal Design}
\label{sec:simulations}
In this section, we assess in different simulations the behaviour of the optimal allocation rate $t_0$ when changing the nominal type-I error rate $\alpha$, the power $1-\beta$, and the ratio of standard deviations $\kappa = \sigma_2/\sigma_1$.

For simulating the influence of $\alpha$, we used $Beta(5,5)$ and $Beta(3,2)$ distributed random numbers in the first and second group. 
For each $\alpha = 0.01, 0.02, \dots, 0.1$, we generated $10^6$ random numbers for each group and calculated the optimal allocation rate $t_0$ and the total sample sizes $N(t_0)$ and $N(1/2)$ 
(corresponding to a balanced design)
to achieve at least $80\%$ power. 
From the formula for the upper bound $I_2$ of $t_0$ we already saw (Section \ref{sec:interval}) that
larger values for the type-I error rate $\alpha$ would lead to a larger difference $|I_2 - 1/2|$. 
While we cannot conclude from this directly that $t_0$ will be more extreme, 
the optimal allocation rate will more likely tend to more extreme values, that is, 
the difference $|t_0 - 1/2|$ tends to become larger. 
We can see this behaviour confirmed in Table \ref{tab:increasing_alpha}. 
In this simulation, we had $p = 0.657$ and $\kappa = 1.53$, implying $t_0 < 1/2$. 
In the data examples, we already found very little difference between using a balanced design or the optimal design. 
The simulation study yielded a similar observation (see Table \ref{tab:increasing_alpha}).

In a second simulation, we investigated the behaviour of $t_0$ for increasing power 
(or decreasing $\beta$). 
We used $\alpha = 0.05$ and the same distributions as before. 
Therefore, $p$ and $\kappa$ were the same as above. 
As power, we chose $1-\beta = 0.6, \dots, 0.95$ and generated $10^6$ random numbers for each $\beta$ to calculate the optimal allocation rate $t_0$. 
The results are displayed in Table \ref{tab:increasing_beta}. 
A larger power led to more extreme values for $t_0$, but the difference in required sample sizes between the balanced and optimal design was again negligible.
\begin{table}[ht]
	\centering
	\begin{tabular}{rrrr}
		\hline
		$t_0$ & $N(t_0)$ & $N(1/2)$ & $\alpha$ \\ 
		\hline
		0.4761 & 153.0998 & 153.4463 & 0.01 \\ 
		0.4742 & 130.2582 & 130.6034 & 0.02 \\ 
		0.4724 & 118.1328 & 118.4890 & 0.03 \\ 
		0.4715 & 108.4745 & 108.8243 & 0.04 \\ 
		0.4704 & 102.7568 & 103.1146 & 0.05 \\ 
		0.4695 & 96.3878 & 96.7427 & 0.06 \\ 
		0.4687 & 91.8895 & 92.2473 & 0.07 \\ 
		0.4680 & 87.5307 & 87.8868 & 0.08 \\ 
		0.4673 & 82.7874 & 83.1389 & 0.09 \\ 
		0.4668 & 79.3812 & 79.7288 & 0.10 \\ 
		\hline
	\end{tabular}
	\caption{Optimal allocation rate $t_0$ and required total sample sizes for optimal allocation and balanced allocation. Power fixed at 80\%. Underlying distributions $Beta(5,5)$ and $Beta(3,2)$.}
	\label{tab:increasing_alpha}
\end{table}
\begin{table}[ht]
	\centering
	\begin{tabular}{rrrr}
		\hline
		$t_0$ & $N(t_0)$ & $N(1/2)$ & $1-\beta$  \\ 
		\hline
		0.4890 & 65.2151 & 65.2464 & 0.60 \\ 
		0.4842 & 72.2678 & 72.3398 & 0.65 \\ 
		0.4793 & 81.0607 & 81.1986 & 0.70 \\ 
		0.4750 & 90.1969 & 90.4212 & 0.75 \\ 
		0.4704 & 102.7568 & 103.1146 & 0.80 \\ 
		0.4658 & 116.2108 & 116.7498 & 0.85 \\ 
		0.4606 & 135.7222 & 136.5547 & 0.90 \\ 
		0.4544 & 166.2805 & 167.6483 & 0.95 \\ 
		
		\hline
	\end{tabular}
	\caption{Behaviour of the optimal allocation rate $t_0$ for increasing power $1-\beta$ with comparison of the sample sizes $N(t_0)$ and $N(1/2)$.}
	\label{tab:increasing_beta}
\end{table}


\section{Discussion}
In this paper, we propose a unified approach to sample size determination for the Wilcoxon-Mann-Whitney two sample rank sum test.
Our approach does not assume any specific type of data or a specific alternative hypothesis. 
In particular, data distributions may be discrete, or continuous.
Based on the general formula, we have also derived an optimal allocation rate to both groups, that is, 
$t = n_1/N$ such that $N$ is minimized. 
The value of this optimal allocation rate $t_0$ mainly depends on the ratio $\kappa = \sigma_2/\sigma_1$
(see (\ref{eq:sig1}) and (\ref{eq:sig2}) for a definition of these variances)
and on $\beta$.
The variance under the null hypothesis has no influence on $t_0$. 
For $\kappa > 1$ we have $t_0 < 1/2$, for $\kappa < 1$ we have $t_0 > 1/2$, and for $\kappa = 1$ we have exactly $t_0 = 1/2$ assuming $u_{1-\beta} > 0$. 
The nominal type-I error rate $\alpha$ only has a small impact on the value of $t_0$. The larger $\alpha$ is, the larger is the difference $|t_0-1/2|$.

We can see from the interval $[I_1, I_2]$ for the optimal allocation rate $t_0$ derived in Section~\ref{sec:interval} 
that $t_0$ will typically be close to $1/2$. 
This was also confirmed in some illustrative data examples in Section \ref{sec:data_results}. 
Furthermore, the difference in required sample size between using a balanced design and using the optimal allocation design appears practically negligible. 
In other words, in most cases, a balanced design can be recommended for the Wilcoxon-Mann-Whitney test.
In extensive simulations, we have confirmed that the new procedure actually meets the power at the calculated sample sizes quite well. 
The new procedure has been implemented into the R package WMWssp and will also be available through the package rankFD \cite{rankFD}.


\section{Acknowledgements} The research was supported by Austrian Science Fund (FWF) I 2697-N31.

\bibliographystyle{plain}

\begin{thebibliography}{10}
	
	\bibitem{akritas1997}
	Michael~G Akritas, Steven~F Arnold, and Edgar Brunner.
	\newblock Nonparametric hypotheses and rank statistics for unbalanced factorial
	designs.
	\newblock {\em Journal of the American Statistical Association},
	92(437):258--265, 1997.
	
	\bibitem{akritas1997b}
	Michael~G Akritas and Edgar Brunner.
	\newblock A unified approach to rank tests for mixed models.
	\newblock {\em Journal of Statistical Planning and Inference}, 61(2):249--277,
	1997.
	
	\bibitem{brunner2000}
	Edgar Brunner and Ullrich Munzel.
	\newblock The nonparametric behrens-fisher problem: asymptotic theory and a
	small-sample approximation.
	\newblock {\em Biometrical Journal}, 42(1):17--25, 2000.
	
	\bibitem{brunner2001}
	Edgar Brunner and Madan~L Puri.
	\newblock Nonparametric methods in factorial designs.
	\newblock {\em Statistical Papers}, 42(1):1--52, 2001.
	
	\bibitem{burkner2017}
	Paul-Christian B{\"u}rkner, Philipp Doebler, and Heinz Holling.
	\newblock Optimal design of the wilcoxon--mann--whitney-test.
	\newblock {\em Biometrical Journal}, 59(1):25--40, 2017.
	
	\bibitem{chakraborti2006}
	Subhabrata Chakraborti, B~Hong, and Mark~A van~de Wiel.
	\newblock A note on sample size determination for a nonparametric test of
	location.
	\newblock {\em Technometrics}, 48(1):88--94, 2006.
	
	\bibitem{collings1988}
	Bruce~Jay Collings and Martin~A Hamilton.
	\newblock Estimating the power of the two-sample wilcoxon test for location
	shift.
	\newblock {\em Biometrics}, pages 847--860, 1988.
	
	\bibitem{dette2004}
	Holger Dette and Timothy~E O'Brien.
	\newblock Efficient experimental design for the behrens-fisher problem with
	application to bioassay.
	\newblock {\em The American Statistician}, 58(2):138--143, 2004.
	
	\bibitem{fan2012}
	Chunpeng Fan and Donghui Zhang.
	\newblock A note on power and sample size calculations for the kruskal--wallis
	test for ordered categorical data.
	\newblock {\em Journal of Biopharmaceutical Statistics}, 22(6):1162--1173,
	2012.
	
	\bibitem{hamilton1991}
	Martin~A Hamilton and Bruce~Jay Collings.
	\newblock Determining the appropriate sample size for nonparametric tests for
	location shift.
	\newblock {\em Technometrics}, 33(3):327--337, 1991.
	
	\bibitem{hilton1993}
	Joan~F Hilton and Cyrus~R Mehta.
	\newblock Power and sample size calculations for exact conditional tests with
	ordered categorical data.
	\newblock {\em Biometrics}, 49(2):609--616, 1993.
	
	\bibitem{kolassa1995}
	John~E Kolassa.
	\newblock A comparison of size and power calculations for the wilcoxon
	statistic for ordered categorical data.
	\newblock {\em Statistics in Medicine}, 14(14):1577--1581, 1995.
	
	\bibitem{rankFD}
	Frank Konietschke, Sarah Friedrich, Edgar Brunner, and Markus Pauly.
	\newblock {\em rankFD: Rank-Based Tests for General Factorial Designs}, 2016.
	\newblock R package version 0.0.1.
	
	\bibitem{lachin2011}
	John~M Lachin.
	\newblock Power and sample size evaluation for the cochran--mantel--haenszel
	mean score (wilcoxon rank sum) test and the cochran--armitage test for trend.
	\newblock {\em Statistics in Medicine}, 30(25):3057--3066, 2011.
	
	\bibitem{leppik1985}
	Ilo~E Leppik, Fritz~E Dreifuss, Terri Bowman, Nancy Santilli, Margaret Jacobs,
	Coral Crosby, James Cloyd, Judy Stockman, Nina Graves, Tom Sutula, et~al.
	\newblock A double-blind crossover evaluation of progabide in partial seizures.
	\newblock {\em Neurology}, 35(4):285, 1985.
	
	\bibitem{lesaffre1993}
	Emmanuel Lesaffre, Ilse Scheys, J{\"u}rgen Fr{\"o}hlich, and Erich Bluhmki.
	\newblock Calculation of power and sample size with bounded outcome scores.
	\newblock {\em Statistics in Medicine}, 12(11):1063--1078, 1993.
	
	\bibitem{matsouaka2016}
	Roland~A Matsouaka, Aneesh~B Singhal, and Rebecca~A Betensky.
	\newblock An optimal wilcoxon--mann--whitney test of mortality and a continuous
	outcome.
	\newblock {\em Statistical Methods in Medical Research}, 0(0):0962280216680524,
	2016.
	\newblock PMID: 27920364.
	
	\bibitem{noether1987}
	Gottfried~E Noether.
	\newblock Sample size determination for some common nonparametric tests.
	\newblock {\em Journal of the American Statistical Association},
	82(398):645--647, 1987.
	
	\bibitem{orban1980}
	John Orban and Douglas~A Wolfe.
	\newblock Distribution-free partially sequential piacment procedures.
	\newblock {\em Communications in Statistics-Theory and Methods}, 9(9):883--904,
	1980.
	
	\bibitem{orban198}
	John Orban and Douglas~A Wolfe.
	\newblock A class of distribution-free two-sample tests based on placements.
	\newblock {\em Journal of the American Statistical Association},
	77(379):666--672, 1982.
	
	\bibitem{puntanen2011}
	Simo Puntanen, George~PH Styan, and Jarkko Isotalo.
	\newblock {\em Matrix tricks for linear statistical models: our personal top
		twenty}.
	\newblock Springer Science \& Business Media, 2011.
	
	\bibitem{rahardja2009}
	Dewi Rahardja, Yan~D Zhao, and Yongming Qu.
	\newblock Sample size determinations for the wilcoxon--mann--whitney test: A
	comprehensive review.
	\newblock {\em Statistics in Biopharmaceutical Research}, 1(3):317--322, 2009.
	
	\bibitem{rosner2009}
	B~Rosner and RJ~Glynn.
	\newblock Power and sample size estimation for the wilcoxon rank sum test with
	application to comparisons of c statistics from alternative prediction
	models.
	\newblock {\em Biometrics}, 65(1):188--197, 2009.
	
	\bibitem{ruymgaart1980}
	Frits~H Ruymgaart.
	\newblock A unified approach to the asymptotic distribution theory of certain
	midrank statistics.
	\newblock In {\em Statistique non Parametrique Asymptotique}, pages 1--18.
	Springer, 1980.
	
	\bibitem{seber2008}
	George~AF Seber.
	\newblock {\em A Matrix Handbook for Statisticians}.
	\newblock John Wiley \& Sons, 2008.
	
	\bibitem{shieh2006}
	Gwowen Shieh, Show-li Jan, and Ronald~H Randles.
	\newblock On power and sample size determinations for the
	wilcoxon--mann--whitney test.
	\newblock {\em Journal of Nonparametric Statistics}, 18(1):33--43, 2006.
	
	\bibitem{tang2011}
	Yongqiang Tang.
	\newblock Size and power estimation for the wilcoxon--mann--whitney test for
	ordered categorical data.
	\newblock {\em Statistics in Medicine}, 30(29):3461--3470, 2011.
	
	\bibitem{thall1990}
	Peter~F Thall and Stephen~C Vail.
	\newblock Some covariance models for longitudinal count data with
	overdispersion.
	\newblock {\em Biometrics}, pages 657--671, 1990.
	
	\bibitem{wang2003}
	Hansheng Wang, Bin Chen, and Shein-Chung Chow.
	\newblock Sample size determination based on rank tests in clinical trials.
	\newblock {\em Journal of Biopharmaceutical Statistics}, 13(4):735--751, 2003.
	
	\bibitem{whitehead1993}
	John Whitehead.
	\newblock Sample size calculations for ordered categorical data.
	\newblock {\em Statistics in Medicine}, 12(24):2257--2271, 1993.
	
	\bibitem{zhao2008}
	Yan~D Zhao, Dewi Rahardja, and Yongming Qu.
	\newblock Sample size calculation for the wilcoxon--mann--whitney test
	adjusting for ties.
	\newblock {\em Statistics in Medicine}, 27(3):462--468, 2008.
	
\end{thebibliography}

\appendix


\section{R Code}
\subsection{Power Simulation}
\begin{verbatim}
x1 # vector of synthetic data of first group
x2 # vector of synthetic data of second group

R <- 10^4
reject <- 0
n1 <- 299
n2 <- 299
set.seed(0)
for(i in 1:R){
z1 <- sample(x1, size = n1, prob = NULL, replace = TRUE)
z2 <- sample(x2, size = n2, prob = NULL, replace = TRUE)

df = data.frame(grp = c(rep(1,n1), rep(2,n2)), z = c(z1,z2))
df$grp <- as.factor(df$grp)

p <- rank.two.samples(z~grp, data = df, wilcoxon = "asymptotic", info = FALSE, shift.int=FALSE, 
alternative = "two.sided")$Wilcoxon$p.Value
if(p <= 0.05){
reject <- reject + 1
}
}
\end{verbatim}

\subsection{Minimize $t$}
\begin{verbatim}
x1 # vector of synthetic data of first group
x2 # vector of synthetic data of second group
alpha = 0.05
beta=0.8
m1 <- length(x1)
m2 <- length(x2)

# ranks among union of samples:
R <- rank(c(x1,x2), ties.method="average")
R1 <- R[1:m1]
R2 <- R[m1+(1:m2)]

# ranks within samples:
R11 <- rank(x1, ties.method="average")
R22 <- rank(x2, ties.method="average")

# placements:
P1 <- R1 - R11
P2 <- R2 - R22

# effect size:
pStar <- (mean(R2)-mean(R1)) / (m1+m2) + 0.5

# variances:
sigmaStar <- sqrt(sum((R11-((m1+1)/2))^2) / m1^3)
sigma1Star <- sqrt(sum((P1-mean(P1))^2) / (m1*m2^2))
sigma2Star <- sqrt(sum((P2-mean(P2))^2) / (m1^2*m2))

sigmaStar <- sqrt(sum( (R- (m1+m2+1)/2)^2  )/(m1+m2)^3)

ss = function(t){
return((sigmaStar*qnorm(1-alpha/2) + qnorm(beta)*sqrt(t*sigma2Star^2 + 
(1-t)*sigma1Star^2))^2 / (t*(1-t)*(pStar-0.5)^2))
}

# sample size with balanced groups
ss(1/2) 

# optimal t
optimize(ss,interval=c(0,1), maximum=FALSE,tol = .Machine$double.eps)$minimum 

# sample size given optimal t
optimize(ss,interval=c(0,1), maximum=FALSE,tol = .Machine$double.eps)$objective

\end{verbatim}


\section{Derivation of the Results}
\subsection{Interval for the Optimal Design}
\begin{satz}
	If we assume $\sigma_1 = \sigma_2$ and $1-\beta > 0.5$  then the optimal design is given by $t_0 = \tfrac{1}{2}$. 
	It is not necessary to assume $1-\beta > 0.5$ but it is convenient to do so in order to avoid a situation where $N(t) = 0$ for all $t \in (0,1)$.
\end{satz}
\begin{proof}
	The numerator of $N(t)$ does not depend on $t$ in this case, therefore $N(t)$ is minimized by $t_0 = \tfrac{1}{2}$.
\end{proof}

\begin{satz}\label{app:result2}
	For $1-\beta > 0.5$ and $0 < \sigma_1 < \sigma_2$ the sample size is minimized by $t_0 \in [I_1, I_2]$ with $I_1 \leq I_2 <
	\tfrac{1}{2}$. The minimizer is unique in the interval $(0,1)$. The bounds $I_1$ and $I_2$ are given by
	\begin{align}
		I_1 &= \frac{1}{\kappa + 1}, \\
		I_2 &= \frac{\sqrt{z}}{ \sqrt{z} + \left(u_{1-\alpha/2} \sqrt{q} \sigma + u_{1-\beta} \sigma_2^2 \right)  },
	\end{align}
	with $\kappa = \sigma_2 / \sigma_1$, $q = p(1-p)$ and $z = \left(u_{1-\alpha/2} \sqrt{q} \sigma + u_{1-\beta} \sigma_1^{2} 
	\right) \left(u_{1-\alpha/2} \sqrt{q} \sigma + u_{1-\beta} \sigma_2^2 \right)$. 
	Additionally the following equivalence holds
	\begin{align}
		t_0 < \tfrac{1}{2} \Longleftrightarrow \sigma_1 <  \sigma_2.
	\end{align}
\end{satz}
\begin{proof}
	First we calculate the derivative of $N$ which is given by
	\begin{align}
		\frac{d}{dt}N(t) = \big(u_{1-\alpha/2}\, \sigma + u_\beta\, \sqrt{\sigma_1^2 \, (1-t) + \sigma_2^2 \,t } \big) \frac{g(t)}{f(t)},
	\end{align}
	where the functions $f$ and $g$ are defined by
	\begin{align*}
		g(t) &= u_{1-\alpha/2}\, \sigma (2t-1) \sqrt{\sigma_1^2\, (1-t) + \sigma_2^2 \,t} - u_\beta\, \big( \sigma_1^2 \, (1-t)^2 - \sigma_2^2 \, t^2 \big),\\
		f(t) &= (p-\tfrac{1}{2})^2 (1-t)^2 t^2 \sqrt{\sigma_1^2 (1-t) + \sigma_2^2 t^2}.
	\end{align*}
	Only $g(t)$ has a root in $(0,1)$. Therefore, we only need to consider this function for finding the optimal $t_0$. To prove the equivalence we start with $t_0 < \tfrac{1}{2}$. 
	In this case, $t_0 > \lambda = \frac{1}{\kappa + 1}$. Because $\tfrac{1}{2} > t_0 > \lambda$ it follows that $\kappa > 1$. The other direction can be proved in a similar manner.
	
	Now that we know $t_0 < \tfrac{1}{2}$ we can easily construct an interval for $t_0$. A lower bound is given by $\lambda$. 
	For the upper bound we use the monotonic function
	\begin{align}
		h(t) = u_{1-\alpha/2}\, \sigma (2t-1) \sqrt{q} - u_\beta \left( \sigma_1^{2} (1-t)^2 - \sigma_2^{2} t^2  \right).
	\end{align}
	This function satisfies $h(t) < g(t)$ for all $t \in (0, \tfrac{1}{2})$ and it has exactly one root $I_2$ in $(0, \tfrac{1}{2})$. From this it immediately follows that $t_0 < I_2$.
	
	For the uniqueness in $(0,1)$, consider a second solution $t_0' \leq t_0$. It follows immediately that $t_0' > \lambda$ and consequently $ \lambda \leq t_0' \leq t_0 \leq \frac{1}{2} $. But $g$ is strictly monotone in $(0, \frac{1}{2})$, therefore both roots are equal.
\end{proof}

\begin{satz}\label{app:result3}
	For $1-\beta > 0.5$ and $\sigma_1 > \sigma_2 > 0$ the sample size is minimized by $t_0 \in [I_2, I_1]$ with $I_1 \geq I_2 > \tfrac{1}{2}$. The minimizer is unique in the interval $(0,1)$. The bounds are the same as in the previous theorem.
	Additionally the following equivalence holds
	\begin{align}
		t_0 > \tfrac{1}{2} \Longleftrightarrow \sigma_1 >  \sigma_2.
	\end{align}
\end{satz}
\begin{proof}
	Similar proof as in the case $0<\sigma_1 < \sigma_2$.
\end{proof}

\begin{satz}
	\label{app:result4}
	For the case $\sigma_1 = 0 < \sigma_2$, we cannot apply the result from before. But using a similar idea we can find a lower bound $l(t)$ for the function $g(t)$ which is defined by
	\begin{align}
		l(t) &= u_{1-\alpha/2}\, \sigma (2t-1)  \sigma_2 \,t + u_\beta\,  \sigma_2^2 \, t^2
	\end{align}
	and this function only has one root  in $(0,1)$, namely
	\begin{align}
		I_1^{(0)} &= \frac{u_{1-\alpha/2}~\sigma }{2 u_{1-\alpha/2}~\sigma + u_{1-\beta}~\sigma_2 } \ = \ \frac1{2 + \gamma} ,
	\end{align}
	where $\gamma = u_{1-\beta}~\sigma_2 \big/ \left( u_{1-\alpha/2}~\sigma \right)$.  
	Then an interval for the optimal design is given by $[I_1^{(0)}, I_2 ]$.
	
\end{satz}

\subsection{Optimality of a Balanced Design}
From the construction of an interval for $t_0$ it is clear that $t_0 = 1/2$ if and only if $\sigma_1^2 = \sigma_2^2$. The equality of variances simply means
\begin{align}
	\int F_2^2 dF_1 - \Bigg(\int F_2 dF_1\Bigg)^2 = \int F_1^2 dF_2 - \Bigg(\int F_1 dF_2\Bigg)^2.
\end{align}
From that we can easily conclude the equivalence
\begin{align}
	t_0 = \tfrac{1}{2} \Longleftrightarrow \int F_1^2 dF_2 = \int (1- F_2)^2 dF_1.
\end{align} 
\begin{satz}
	Let us now consider normalized cumulative distribution functions $F_1, F_2$ for which an $a \in \mathbb{R}$ exists such that for all $x \in \mathbb{R}$ Equation (\ref{eq:distributions}) holds, that is,
	\begin{align}
		F_1(a + x ) = 1 - F_2(a - x).
	\end{align}
	Then the optimal design is given by $t_0 = 1/2$. Furthermore if such an $a$ exists and the expectations of the two distributions are finite, then the constant $a$ can be explicitly calculated as
	\begin{align} \label{eq:a}
		a = \frac12 \left( \int x dF_1(x) + \int x dF_2(x) \right),
	\end{align}
	that is, $a$ is the average of the expected values. If the third moments are finite, then it follows from 
	(\ref{eq:distributions}) that the variances of the distributions $F_1$ and $F_2$ are equal and their skewness have opposite sign. In the case $F_1 = F_2$, 
	the assumption (\ref{eq:distributions}) simply means that $F_1$ is a symmetric distribution.
\end{satz}

\begin{proof}
	This equivalence holds since $F_1$ and $F_2$ satisfy $\int F_1^2 dF_2 = \int (1- F_2)^2 dF_1$.
	Equation (\ref{eq:a}) follows directly after some calculations by first considering $F_1$ and $F_2$ to be either continuous or 
	discrete. Then (\ref{eq:a}) also holds for distributions with a continuous and discrete proportion. First we proof (\ref{eq:a}) for 
	the discrete case. Note that from (\ref{eq:distributions}) we can conclude that $P(X_1 = x) = P(X_2 = 2a - x)$ holds. Then 
	for discrete $X_1 \sim F_1$ and $X_2 \sim F_2$ the result follows from
	\begin{align*}
		E X_1 &= \sum_{i} x_i P(X_1 = x_i) \\&= - \sum_{i} (2a - x_i) P(X_2 = 2a - x_i) + 2a \ = \ - E X_2 + 2a \ .
	\end{align*}  
	The derivation for the continuous case is similar.
\end{proof}

\end{document}